\documentclass{emulateapj}
\usepackage{apjfonts}

\newcommand{\beq}{\begin{equation}}
\newcommand{\beqa}{\begin{eqnarray}}
\newcommand{\eeq}{\end{equation}}
\newcommand{\eeqa}{\end{eqnarray}}

\newcommand{\simg}{\ga}
\newcommand{\meszaros}{M\'esz\'aros}

\slugcomment{}

\shorttitle{Hypernova/GRB Remnants as TeV unIDs}
\shortauthors{Ioka \& \meszaros}

\begin{document}

\title{
Hypernova and Gamma-Ray Burst Remnants
as TeV Unidentified Sources
}


\author{Kunihito Ioka\altaffilmark{1}, 
and 
Peter \meszaros\altaffilmark{2}}


\altaffiltext{1}{
%\affil{
%Theory Division, KEK (High Energy Accelerator Research Organization)
KEK Theory Center
and the Graduate University for Advanced Studies (Sokendai), 
1-1 Oho, Tsukuba 305-0801, Japan
}
\altaffiltext{2}{
%\affil{
Center for Particle Astrophysics, 
Dept. of Astronomy and Astrophysics, 
Dept. of Physics, Pennsylvania State University, University
Park, PA 16802
}

%\email{kunihito.ioka@kek.jp}

\begin{abstract}
We investigate hypernova (hyper-energetic supernova) and 
gamma-ray burst (GRB) remnants in our Galaxy as TeV gamma-ray sources, 
particularly in the role of potential TeV unidentified sources, 
which have no clear counterpart at other wavelengths.
We show that the observed bright sources in the TeV sky 
could be dominated by GRB/hypernova remnants,
even though they are fewer than supernova remnants (SNRs).
If this is the case, TeV SNRs are more extended (and more numerous) 
than deduced from current observations. In keeping with their role
as cosmic ray accelerators, we discuss hadronic gamma-ray emission 
from $\pi^0$ decay, from $\beta$ decay followed by inverse Compton 
emission, and propose a third, novel process of TeV gamma-ray emission
arising from the decay of accelerated radioactive isotopes such as 
${}^{56}$Co entrained by relativistic or semi-relativistic jets in 
GRBs/hypernovae. We discuss the relevant observational signatures which 
could discriminate between these three mechanisms.
\end{abstract}


\section{Introduction}

Observations at a new wavelengths have always led in astronomy to 
new discoveries, such as radio pulsars, gamma-ray bursts (GRBs), etc.
The TeV gamma-ray sky is likely to yield similar surprises,
since among the rapidly growing number of TeV sources \citep{aha06}
the most abundant category is of a mysterious nature, the so-called 
TeV unidentified sources (TeV unIDs), which have no clear counterpart 
at other wavelengths \citep{aha05,aha08,reshmi05}.
Our knowledge of TeV unIDs is very limited:
\begin{itemize}
\item[(1)] So far $N_{\rm unID} \sim 10$-$30$ TeV unIDs have been observed.

\item[(2)] They generally lie close to the Galactic plane, suggesting a Galactic origin.

\item[(3)] They are extended, $\Delta\Omega\sim 0.05$-$0.3^{\circ}$.

\item[(4)] The flux is $\varepsilon_\gamma F_{\varepsilon_\gamma} \sim 10^{-12}$-$10^{-11}$
${\rm erg\ s}^{-1}$ ${\rm cm}^{-2}$ at $\varepsilon_\gamma \sim 0.2$ TeV.

\item[(5)] They have a power-law spectrum with index of $2.1$-$2.5$.

\item[(6)] Some TeV unIDs have strong upper limits in X-rays with a TeV to 
X-ray flux ratio of $F_{\rm TeV}/F_{\rm X} \simg 50$ from Suzaku 
\citep{mat07,bam07} and in radio with 
$F_{\rm TeV}/F_{\rm radio} \sim 10^3$ \citep{atoyan06,tian08}.
\end{itemize}

The high energy nature of TeV unIDs naturally leads us to consider 
the cosmic ray (CR) accelerators as their possible origin.
Since the dominant CR sources are supernova remnants (SNRs),
TeV unIDs may be related to SNRs, in particular old ones, which are
expected to be less luminous at other wavelengths \citep{yam06}.
However, the required galactic energy budget of TeV unIDs is only 
\beqa
4 \pi d^2 \varepsilon_\gamma F_{\varepsilon_\gamma} N_{\rm unID}
\sim 10^{34{\rm -}35} \left(\frac{d}{10{\rm kpc}}\right)^{2} {\rm erg\ s}^{-1},
\eeqa
which is much less than that of SNe,
\beqa
\frac{10^{50}{\rm erg}}{100 {\rm yr}} \sim 10^{41} {\rm erg\ s}^{-1},
\eeqa
implying a rarer type of source.

Long GRBs are actually rare SNe endowed with relativistic jets,
which also are expected to leave SNR-like remnants \citep{per00,aya01}.
Even so, the expected number of GRB remnants is so small
that previous works have addressed only specific sources 
\citep{ioka04,atoyan06}.
However, recent observations suggest that hyper-energetic SNe,
the so-called hypernovae which are sometimes 
associated with GRBs such as SN1998bw/GRB980425,
SN2003dh/GRB030329 and SN2003lw/GRB031203
\citep{mn02},
occur more frequently than GRBs \citep{gue07}.
The hypernova rate may be even higher
if we are missing GRB-unassociated hypernovae like SN1997ef.
On the other hand, a larger fraction of SNe may be endowed with
slower or semi-relativistic jets, causing only low-luminosity (LL) GRBs.
Actually, a significant energy $\sim 10^{50}$ erg was released 
in the form of a mildly relativistic ejecta
in the very faint GRB980425/SN1998bw.
The recently discovered LL GRB060218/SN2006aj
could be also be produced by a slower or semi-relativistic jet 
\citep{toma07,waxman07}, occurring at a $\sim 10$ times 
higher rate than GRBs \citep{gue07,lia07,sod06c}.
Such semi-relativistic jets could be intermediate between
choked and breaking-out jets emerging through
the progenitor's outer envelope.

An interesting point in TeV unIDs
is that their emission may be related with hadronic processes,
since a simple leptonic process for TeV gamma-rays
by inverse Compton (IC) would require too much synchrotron 
emission than what is observed in TeV unIDs.
Therefore we consider the hadronic processes associated 
with the $\pi^0$ decay (\S~\ref{sec:pi}) for the hypernova shocks,
and with the $\beta$ decay mechanism (\S~\ref{sec:beta}), 
as well as a new mechanism of accelerated radio-isotope (RI) decay 
(\S~\ref{sec:ri}) for the GRB jets.
In the RI decay model, TeV gamma-rays are produced by the 
Lorentz-boosting of the MeV decay gamma-rays of accelerated RI, 
such as ${}^{56}$Co entrained by the jets.
To our knowledge, this is the first discussion of the
decay of accelerated RI as a mechanism for generating TeV 
gamma-rays in astrophysical sources. In this case, the source
of the gamma-ray energy is injected initially at the base of 
the jet. There is no need for target matter or photons, 
as in the usual processes such as the $\pi^0$ decay,
inverse Compton (IC) or photodisintegration \citep{anc07}.

In this paper, we investigate the high energy implications of 
GRB/hypernova remnants for TeV sources, in particular for TeV unIDs.
Even though the total number of GRB/hypernova remnants is less 
than that of SNRs, the observed number in the TeV sky 
could be larger (\S~\ref{sec:pi}).
This in turn predicts more extended (and more numerous) TeV SNRs 
than currently observed, which may be discovered by expanding 
the search region even with the current instruments.
We use the units $k_B=h=1$
and $Q_x=Q/10^x$ in cgs units unless otherwise stated.

\section{Hypernova shocks}
\subsection{$\pi^0$ decay model}\label{sec:pi}

The simplest and most plausible process for TeV gamma-ray emission associated
with CRs is the $\pi^0$ decay from $pp$ interactions between the 
interstellar medium (ISM) and CRs accelerated by the SN/hypernova shocks.
Although the $\pi^0$ decay for SNRs \citep{nt94,dru94}
and GRB remnants \citep{atoyan06} has been discussed in detail,
there are few works dealing with $\pi^0$ decay in hypernova remnants,
except for the specific source HESS J1303-631 \citep{atoyan06}.
Since hypernovae are more energetic than SNe and GRBs and more frequent 
than GRBs, it is worth considering their implications for TeV unIDs.

By scaling the SNR calculation \citep{nt94,dru94}, 
we obtain a TeV gamma-ray flux
\beqa
\varepsilon_\gamma F_{\varepsilon_\gamma}
\sim 10^{-12} \zeta_{-1} E_{51} n 
d_{10{\rm kpc}}^{-2}\
{\rm erg\ s}^{-1}\ {\rm cm}^{-2},
\label{eq:fpi}
\eeqa
comparable to the observed values. Here $n$ is the ISM density in cm$^{-3}$,
$d_{10{\rm kpc}}=d/10{\rm kpc}$ and $\zeta$ is the fraction of the 
total CR energy $E$ per logarithmic energy interval.
Since the kinetic energy of hypernovae is huge, 
$E_k \sim 10^{52}$ erg, a CR energy of $E \sim 10^{51}$ erg is 
reasonable, for a conventional 
CR acceleration efficiency of $\sim 10\%$.

\begin{figure}
\plotone{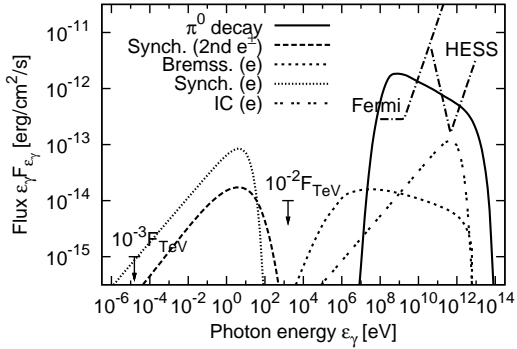}
\caption{
Flux from the $\pi^0$ decay via $pp$ interactions between
the ISM and CRs accelerated by the hypernova shocks,
compared with the Fermi and HESS sensitivities.
We assume a remnant of age $t_{\rm age}=10^5$ yr at $d=10$ kpc
with a CR energy $E=3 \times 10^{51}$ erg 
and the CR spectral index $p=2.2$ in the energy range
$m_p c^2 < \varepsilon_p < 10^5 m_p c^2$.
We also show the synchrotron emission from $\pi^0$ decay 
positrons and electrons for $B=3\mu$G, compared with 
the observational upper limits for an X-ray to TeV 
flux ratio of $10^{-2}$ and a radio to TeV flux ratio of $10^{-3}$.
We use a code of \citet{kamae06} for calculating $pp$ interactions.
In addition, we also plot the bremsstrahlung, synchrotron,
and IC emission from primary electrons with
the spectral index $p=2.2$
and the electron-to-proton ratio of 
$(e/p)_{10{\rm GeV}}\sim m_e/m_p$ at $10$GeV.
}
\label{fig:pi}
\end{figure}

Since the flux in equation (\ref{eq:fpi}) is independent of time,
most sources are likely old, e.g., $t_{\rm age}\sim 10^5$ yr old.
Such old SNRs are possible TeV unIDs because the maximum energy of 
primary electrons is so small that the leptonic emission is suppressed.
As shown in Fig.~\ref{fig:pi},
the synchrotron emission from secondary electrons can also be 
below upper limits for TeV unIDs in the radio to X-ray bands
\citep{yam06,atoyan06}.

Remarkably, the {\it observed} number of old hypernova remnants 
may be comparable to that of the more numerous old SNRs,
if the observations are flux-limited.
For old remnants, which are larger than the angular resolution
[see equation (\ref{eq:size})],
the source size should be taken into account.
As the search region is expanded, more background is included.
The sensitivity is proportional to the inverse square of the background,
for background-dominated counting statistics.
Therefore, the flux sensitivity to an extended source with a physical radius
$r$ is given by $F_{\varepsilon_\gamma}^{\rm extend}
=F_{\varepsilon_\gamma}^{\rm point}(r/d \theta_{\rm cut})$,
where $F_{\varepsilon_\gamma}^{\rm point}$ is the sensitivity to a point source
and $\theta_{\rm cut}$ is the angular cut in the analysis
\citep{kon02,les01}.
With equation (\ref{eq:fpi}) and
\beqa
F_{\varepsilon_\gamma}^{\rm point}\left(\frac{r}{d \theta_{\rm cut}}\right)
<F_{\varepsilon_\gamma} \propto E~ n ~ d^{-2},
\eeqa 
we have the maximum distance to a source as
\beqa
d_{\max} \propto E~ n~ r^{-1}
\eeqa
and hence the observable volume in the Galactic disk is
\beqa
V \propto d_{\max}^2 \propto E^2 n^2 r^{-2} \propto E^{8/5} n^{12/5},
\eeqa
where we use $r \propto E^{1/5} n^{-1/5} t_{\rm age}^{2/5}$, which is 
approximately correct even for the radiative phase of the remnants.
Since the hypernova energy is $\sim 10$ times larger,
the observable number of hypernova remnants is larger than that of SNRs by
\beqa
\frac{N_{\rm HNR}^{\rm obs}}{N_{\rm SNR}^{\rm obs}}
\sim \frac{R_{\rm HNR} V_{\rm HNR}}{R_{\rm SNR} V_{\rm SNR}}
\sim \frac{10^{-4}{\rm yr}^{-1} \cdot 10^{8/5}}
{10^{-2}{\rm yr}^{-1} \cdot 1^{8/5}}
\sim 0.4,
\eeqa
where $R_{\rm SNR} \sim 10^{-2}{\rm yr}^{-1}$ and 
$R_{\rm HNR} \sim 10^{-4}{\rm yr}^{-1}$ ($\sim 7\%$ of the SNe Ibc rate)
are the event rates of SNe and hypernovae, respectively \citep{gue07},
and we assume a similar age $t_{\rm age}$ and 
ISM density $n$ for both remnants.
The actual numbers in our Galaxy would be
$R_{\rm SNR} t_{\rm age} \sim 10^3$ (SNRs) and 
$R_{\rm HNR} t_{\rm age} \sim 10$ (hypernova remnants)
for $t_{\rm age} \sim 10^5$ yr.
Since the observed number of TeV unIDs is also $\sim 10$, 
we may be reaching the farthest hypernova remnants in our Galaxy.

The angular size of the $t_{\rm age} \sim 10^5$ yr old remnants,
\beqa
\frac{r}{d} \sim \frac{30{\rm pc}}{d} 
\sim 0.2^{\circ} \left(\frac{d}{10{\rm kpc}}\right)^{-1},
\label{eq:size}
\eeqa
is consistent with TeV unIDs for hypernovae,\footnote{
The CR diffusion could increase the source size \citep{atoyan06},
although the diffusion coefficient has large uncertainties.}
 but the same may not
be true for SNe since the observable distance $d_{\max}$ is smaller
and hence the angular size is more extended
[$r/d \sim 2^{\circ} (1{\rm kpc}/d)$].
Such an extended object can be found only if the search region
is expanded to the degree scale in the analyses.
Therefore, if TeV unIDs are hypernova remnants,
we can predict more extended (and more numerous) TeV SNRs than observed,
which may be discovered by expanding the search region
even with the current instruments.

The flux of hypernova remnants could be further enhanced if the
density $n$ is higher around hypernovae than around usual SNe.
The GRBs and SNe Ic, which are the only type of core collapse SNe 
associated with GRBs and hypernovae, are far more concentrated 
in the very brightest regions of their host galaxies 
than are the ohter SNe \citep{kel07,fru06},
suggesting that GRBs and hypernovae are associated with star 
forming regions with high density.
Conversely, hypernovae could be runaway massive stars ejected 
several hundreds of parsec away from high density stellar clusters
\citep{ham06}.
In this case, no flux enhancement is expected.

It is necessary to check any possible side-effects of this mechanism
at other wavelengths which might violate the hypthesis that they
remain unidentified. In particular we need to check what is the 
possible X-ray emission from electrons in these sources.
We may constrain the electron acceleration in the old hypernova 
remnants.  In Fig.~\ref{fig:pi}, we have plotted the bremsstrahlung, 
synchrotron, and IC emission from primary electrons with the same 
spectral index $p=2.2$, for an electron-to-proton ratio of 
$(e/p)_{10{\rm GeV}}\sim m_e/m_p \sim 10^{-3}$ at $10$GeV.
The radio observations limit the $(e/p)_{10{\rm GeV}}$ ratio to
$\lesssim 10^{-3}$, which is somewhat smaller than the observations $\sim 0.02$.
Note however that the injection of thermal electrons into the acceleration 
process is poorly understood, and the observed electron CRs might be produced 
by other sources such as usual SNRs or pulsars.
In any case, the leptonic model for TeV unIDs seems unlikely.
Note that we take the Coulomb losses for low energy electrons
into account according to \citet{uchi02}
(see also \citet{baring99})
and this makes the bremsstrahlung spectrum hard in the X-ray band.
On the other hand, 
the thermal bremsstrahlung emission
is not bright in the X-ray band,
in contrast with the usual young SNRs
\citep{Katz:2007ds},
because the temperature of old SNRs
is much below the X-ray energy,
$T \sim 0.04\ {\rm keV}
\ E_{k,52}^{2/5} n^{-2/5} (t/10^5{\rm yr})^{-6/5}$
\citep{reynolds08}.
%The thermal bremsstrahlung emission is below
%the X-ray band since older remnants have lower temperature.

\section{GRB jets}
\subsection{$\beta$ decay model}\label{sec:beta}

A fraction of SNe/hypernovae is associated with long GRB and their jets.
Another process for generating TeV gamma-rays is the (time-delayed)
$\beta$ decay of the neutron component of the CR outflow accelerated by 
the jets followed by inverse Compton scattering \citep{ioka04}.
We have applied this process to the SNR W49B (G43.3-0.2), 
which is a possible GRB remnant
\citep{keo06,mic06,mic08}.
%Near-infrared images reveal a barrel-shaped structure, while X-ray data show 
%a jet-like structure with an enhancement of heavy elements, particularly Fe/Ni 
%in the core and jet regions. This suggests that the explosion may have been 
%jet driven as expected in GRBs and hypernovae.
Since the $\beta$ decay can occur outside the remnant,
the jet-like emission would appear outside of the SNR.
Note that the GRB-associated SNe are not necessarily energetic hypernovae
since the luminosity distribution of GRB-associated SNe
is statistically consistent with that of local SNe Ibc \citep{sod06b}.

\begin{figure}
\plotone{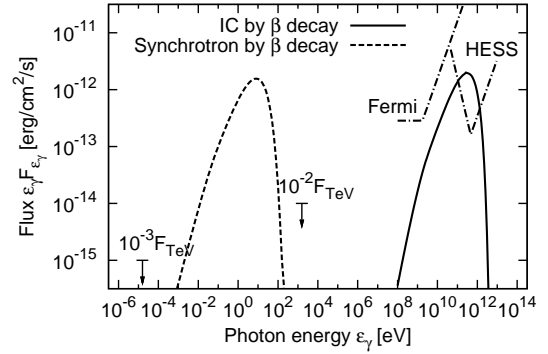}
\caption{
Flux from the IC scattering of CMB photons
by the $\beta$ decay electrons via the CR neutron component in the GRB jets,
compared with the Fermi and HESS sensitivities.
We assume a remnant of age $t_{\rm age}=10^5$ yr at $d=10$ kpc
with a CR energy $E=3 \times 10^{51}$ erg 
[i.e., an old remnant version of Model (I) in \citet{ioka04}]
and a CR Lorentz factor $10^6 < \gamma_n < 10^{9}$.
We also show the synchrotron emission from $\beta$ decay electrons
for $B=3\mu$G,
compared with observational
upper limits for an X-ray to TeV flux ratio of $10^{-2}$
and a radio to TeV flux ratio of $10^{-3}$.
}
\label{fig:beta}
\end{figure}

As shown in Fig.~\ref{fig:beta},
old jet remnants with $t_{\rm age} \sim 10^{5}$ yr are potentially TeV 
unIDs since the usual SNR emission goes down at this age,
while electrons that emit $\sim $TeV gamma-rays 
(i.e., $\gamma_e \sim 10^{7}$ electrons)
need $t_{\rm cool} \sim 10^{5}$ yr to cool
\citep[see Eq.~(3) and (4) in][]{ioka04}.
(However a spectral cutoff at $\sim 10$ TeV may be seen at this age.)
Then the expected total number 
is $0.1$-$1$ for a Galactic GRB rate of $\sim 10^{-5}$-$10^{-6}$ yr$^{-1}$
\citep{gue07}.
The actual number would be larger because
a larger fraction of SNe may have lower-luminosity (LL) jets
that were not identified \citep[e.g.,][]{toma07},
and LL jets can also accelerate CRs
to energy $\sim 10^{16{\rm -}17}$ eV
\citep[e.g.,][]{minn06,minn08},
which is necessary for the TeV gamma-ray emission.
The most optimistic rate consistent with the late-time radio observations
is $\sim 10^{-4}$ yr$^{-1}$ ($\sim 10\%$ of the SN Ibc rate) \citep{sod06a}, 
yielding a total number of $\sim 10$, comparable to that of TeV unIDs.
All of these remnants could be detected,
i.e., $d_{\rm max}\sim 10$kpc, if the total energy in 
accelerated CRs is comparable to the typical GRB energy.

\subsection{Radio-isotope (RI) decay model}\label{sec:ri}
%\subsection{Energetics of radioactive jets}

The chemical composition of jets which may be associated with
SNe/hypernovae is currently unknown. One interesting possibility is 
that they entrain radioactive isotopes (RI), 
in particular ${}^{56}$Ni and ${}^{56}$Co.
The SN/hypernova shock leads to explosive nucleosynthesis, predominantly 
forming ${}^{56}$Ni via complete silicon burning \citep{mn03}.
The inner ${}^{56}$Ni would fall back onto the central engine
(in those objects where there is one), and a fraction of 
accreted ${}^{56}$Ni could be ejected with the ensuing jet.
The photodisintegration of heavy nuclei would be suppressed 
in the LL jets whose temperature is less than $\sim$MeV.
Even if heavy nuclei may disintegrate in the innermost accretion disk,
${}^{56}$Ni could be produced again in a cooling wind \citep{mac03}.
${}^{56}$Ni could be also entrained into the jet from the surroundings.
The subsequent photodisintegration can be avoided
in large fraction of parameter space \citep{wrm08,minn08}.
The spectrum and light curve modeling of GRB-associated hypernovae suggests 
that the total amount of ${}^{56}$Ni is larger than that in ordinary SNe
and heavy elements are aspherically ejected along the jet direction
\citep{mn02}.

${}^{56}$Ni decays into ${}^{56}$Co in $t_{\rm Ni}\ln 2=6.1$ days 
essentially by electron capture \citep{mw02}.
The decay proceeds primarily through the excited states of ${}^{56}$Co,
emitting $\sim 2$ MeV gamma-rays.
${}^{56}$Co also decays primarily by electron capture into excited ${}^{56}$Fe 
with a half-life of $t_{\rm Co}\ln 2=77.2$ days,
\beqa
{}^{56}{\rm Co} + e^{-} \to {}^{56}{\rm Fe}^{*} + \nu_e,
\eeqa
followed by $\varepsilon_{\rm RI} \sim 2$ MeV gamma-ray emission,
\beqa
{}^{56}{\rm Fe}^{*} \to {}^{56}{\rm Fe} + \gamma.
\label{eq:Fe*}
\eeqa
If the nuclei are fully ionized,
they decay by emission of a positron ($\beta^{+}$ decay), e.g.,
\beqa
{}^{56}{\rm Co} \to {}^{56}{\rm Fe}^{*} + e^{+} + \nu_e,
\label{eq:Cob+}
\eeqa
followed by gamma-ray emission,
since free electron capture is negligible in our case.
The half-life for ionized ${}^{56}$Co
is a factor of $\sim 5$ higher than for ${}^{56}$Co with electrons,
while that of ionized ${}^{56}$Ni
is longer than $3 \times 10^4$ yr.
Since nuclei can recombine in an expanding jet \citep{mw02},
the decay chain ${}^{56}{\rm Ni} \to {}^{56}{\rm Co} \to
{}^{56}{\rm Fe}$
can start from the ${}^{56}$Ni electron capture.

The outflowing jet is shocked as it interacts with the ISM, 
leading to particle acceleration. Unlike in the ordinary SNe,
the RI acceleration can occur before the RI decay in the objects
considered here, involving jets reaching relativistic speeds.
Our interest is in the ${}^{56}$Co acceleration,
since the accelerated nuclei are completely ionized
and the ionized ${}^{56}$Ni is essentially stable
(the half-life is longer than $3\times 10^4$ yr).
For the ${}^{56}$Co acceleration,
the shock radius should be smaller than the ${}^{56}$Co decay length,
$r<c t_{\rm Co} (\ln 2) \beta \Gamma$
and larger than the ${}^{56}$Ni one,
$r>c t_{\rm Ni} (\ln 2) \beta \Gamma$,
where $\Gamma$ is the Lorentz factor of the jet.
Since the shock radius may be estimated by
$E_j \sim (4\pi/3) r^3 n m_p c^2 \beta^2 \Gamma^2$,
we have
\beqa
2 E_{j,51}^{1/5} n^{-1/5} < \beta \Gamma < 8 E_{j,51}^{1/5} n^{-1/5}
\eeqa
where $E_j$ is the total energy of the jet.
The Lorentz factor ranges between that of GRBs and hypernovae.
Note that the shock radius is relatively large $\simg 10^{16}$ cm
that no heavy nuclei may disintegrate \citep{minn08,wrm08}.

Once ${}^{56}$Co is accelerated, the observed mean lifetime is extended 
by the Lorentz factor, %of the ${}^{56}$Co,
\beqa
t_{\gamma}=\gamma t_{\rm iCo} \sim 10^5 {\rm yr}\,
\gamma_{5}
\label{eq:tga}
\eeqa
enabling long-lasting emission, where $t_{\rm iCo}=5 t_{\rm Co}$
is the half-life of ionized ${}^{56}$Co.
The observed energy of the decay gamma-rays is also Lorentz-boosted to
\beqa
\varepsilon_{\gamma}=\gamma \varepsilon_{\rm RI} 
\sim 0.2 {\rm TeV} \gamma_{5},
\eeqa
potentially applicable to TeV unIDs.
The ratio of gamma-ray energy to the ${}^{56}$Co CR energy is about
$f={\varepsilon_{RI}}/56 m_p c^2
\sim 4 \times 10^{-5}$.
The gamma-ray flux is then estimated as
\beqa
\varepsilon_\gamma F_{\varepsilon_\gamma} 
=\frac{f \zeta E}{4\pi d^2 t_\gamma}
\sim 10^{-12} {\rm erg}\ {\rm s}^{-1}\ {\rm cm}^{-2}
\frac{\zeta_{-1} E_{51}}{\gamma_{5} d_{3{\rm kpc}}^{2}},
\label{eq:lumi}
\eeqa
interestingly comparable to that of TeV unIDs,
where $d_{3{\rm kpc}}=d/3{\rm kpc}$ and
$E$ is the total energy of ${}^{56}$Co CRs.
In a sense, this is a relativistic version of the SN emission
since the energy source is the RI decay.

In the RI model, a similar way to the $\beta$ decay model, 
the total source number is $0.1$-$1$ for a Galactic GRB rate of 
$\sim 10^{-5}$-$10^{-6}$ yr$^{-1}$ \citep{gue07},
while the most optimistic number is $\sim 10$
for a rate consistent with the late-time radio observations
$\sim 10^{-4}$ yr$^{-1}$ ($\sim 10\%$ of the SN Ibc rate) \citep{sod06a}.
We can detect a fraction $\sim (3{\rm kpc}/10{\rm kpc})^2 \sim 0.1$
of these sources, if CRs comprise an energy comparable to the typical GRB energy.
For a rare event with high energy and/or
with reacceleration by the hypernova shock,
the observable distance may be farther.

We note that other RI species may also contribute.
In particular, the ${}^{57}{\rm Ni}$ emission via $\beta^+$ decay
dominates in the young ($\sim 10^3$ yr) remnants because
its half-life $35.6$ hr is $\sim 10^2$ times shorter than ${}^{56}{\rm Ni}$
though its yield is less by $0.1$-$0.01$ \citep{mn02,mn03}.

%\subsection{Broadband spectrum}

The shocked ${}^{56}$Co is accelerated to a power-law spectrum
$dN_{\rm Co} \propto \varepsilon_{\rm Co}^{-p} d \varepsilon_{\rm Co}$
for $\varepsilon_{\rm Co}<\varepsilon_{\rm Co, max}$.
The maximum energy can be
$\varepsilon_{\rm Co, max} \sim 3 \times 10^{17}$ eV 
\citep[e.g.,][]{minn08},
which provides
$\varepsilon_{\gamma,\max} \sim f\times (3 \times 10^{17}{\rm eV}) \sim 10$ TeV 
decay gamma-ray.
Hereafter we adopt this maximum energy.
The Larmor radius of ${}^{56}$Co CRs is
$\sim 4 (\varepsilon_{\rm Co}/10^{17}{\rm eV}) B_{-6}^{-1}$ pc,
leading to the isotropic emission of decay gamma-rays.

\begin{figure}
\plotone{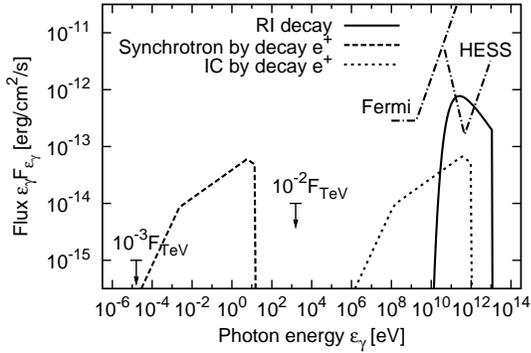}
\caption{
Flux of the decay gamma-rays from accelerated RI ${}^{56}$Co CRs
in equations (\ref{eq:Fe*}), (\ref{eq:lumi}) and (\ref{eq:gspec}),
compared with the Fermi and HESS sensitivities.
We assume a remnant of age $t_{\rm age}=10^5$ yr at $d=3$ kpc
with a CR energy $E=3 \times 10^{51}$ erg and spectral index $p=1.5$.
We also show the synchrotron and IC emission from decay positrons
given by equation (\ref{eq:Cob+}) for $B=3\mu$G
and the cosmic microwave background,
compared with observational
upper limits for an X-ray to TeV flux ratio of $10^{-2}$
and a radio to TeV flux ratio of $10^{-3}$.
}
\label{fig:flux}
\end{figure}

As shown in Fig.~\ref{fig:flux}, the decay gamma-rays have a unique spectrum,
\beqa
\varepsilon_\gamma F_{\varepsilon_\gamma} 
\propto \varepsilon_\gamma^{-p+1}
\exp\left(-\frac{{\varepsilon_{p}}}{{\varepsilon_\gamma}}\right)
\label{eq:gspec}
\eeqa
with an exponential cutoff \footnote{
The actual cutoff is a power low 
$\varepsilon_\gamma F_{\varepsilon_\gamma} \propto \varepsilon_\gamma^{2}$
for $\varepsilon_{\gamma}<\varepsilon_{p}$
because nuclei also make off-axis emission outside the relativistic beaming angle
\citep{ioka01}. This does not change our results.
} 
below 
\beqa
\varepsilon_p=\varepsilon_{\rm IR} \left(\frac{t}{t_{\rm iCo}}\right)
\sim 0.2 {\rm TeV} \left(\frac{t}{10^5{\rm yr}}\right)
\label{eq:epeak}
\eeqa
and a power-law above it, where 
the power-law index is softer than that of parent CRs by one.
This is because the low energy CRs have already decayed
while the high energy CRs are decaying with a rate
$(d/dt)\exp(-t/\gamma t_{\rm iCo})\propto 
\varepsilon_\gamma^{-1} \exp(-\varepsilon_p/\varepsilon_\gamma)$.
%The peak energy $\varepsilon_p$ moves to the high energy as time increases.
The spectrum of TeV unIDs is characterized as a
power-law with index $2.1$-$2.5$ \citep{aha08}, which corresponds 
to $p=1.1$-$1.5$ in equation (\ref{eq:gspec}).
This is slightly harder than the $p=2$ 
expected in the test particle acceleration case.
The harder spectrum could suggest non-linear shock effects in jets 
where the CR pressure is nonnegligible \citep{ell01}.
%\citep{ell01,md01}.
This is consistent with the fact that the necessary CR energy 
is comparable to the total energy of jets, $E \sim E_j$.

In Figure~\ref{fig:flux}, we also show the 
synchrotron and IC emission from 
decay positrons in equation (\ref{eq:Cob+}), 
which have an energy comparable to that of decay gamma-rays.
The positron spectrum is $dN_e \propto \varepsilon_e^{-p} d\varepsilon_e$
for $\Gamma m_e c^2< \varepsilon_e < \varepsilon_p$
and $dN_e \propto \varepsilon_e^{-p-1} d\varepsilon_e$
for $\varepsilon_p < \varepsilon_e < \varepsilon_{\max}$.
The synchrotron frequency 
is $\nu^{\rm syn}=q B \varepsilon_e^2/2\pi m_e^3 c^5
\sim 10^{-4} B_{-6} (\varepsilon_e/0.1{\rm TeV})^2$ eV
while the IC frequency is
$\nu^{\rm IC} \sim 9 \epsilon_{\rm CMB} (\varepsilon_e/m_e c^2)^2
\sim 10^8 (\varepsilon_e/0.1{\rm TeV})^2$ eV
where we consider cosmic microwave background (CMB) as main target photons
\citep{ioka04}.

However the cooling time 
$t_c = 3 m_e^2 c^3/4 \sigma_T \varepsilon_{e} U
\sim 10^7 (\varepsilon_e/0.1{\rm TeV})^{-1} U_{-12}^{-1}$ yr
is usually longer than the decay time $t_\gamma$ in equation (\ref{eq:tga}) for 
$\varepsilon_e < \varepsilon_{\max} \sim 10$ TeV
where $U$ is the total energy density of 
magnetic fields and CMB.
Then the luminosity is suppressed by
$t_\gamma/t_c \sim 10^{-2} (t_\gamma/10^5{\rm yr}) 
(\varepsilon_e/0.1{\rm TeV}) U_{-12}$
compared with decay gamma-rays in equation (\ref{eq:lumi}).
This is favorable for the interpretation of such sources
as TeV unIDs.  Note that the $\sim$eV photons by decay positrons
come from an extended region in which the optical background
dominates the remnant flux.

The accelerated RI CRs may suffer the adiabatic energy losses
during the expansion of the remnant \citep{Rachen:1998fd}, 
which will greatly suppress the RI emission.
This is a potential problem for the RI decay model.
However the CR escape from a remnant is not fully understood yet.
For instance, it is common to assume the free escape of CRs 
in models where the GRB reverse shocks are the sites for the ultra 
high energy CR production \citep[e.g.,][]{Waxman:2004ez,minn08}.
In the same sense, the RI decay model is a viable possibility for TeV unIDs.

\section{Discussion}

We have discussed the TeV gamma-ray emission from the $\pi^0$ decay, 
$\beta$ decay and the radio-isotope (RI) decay mechanisms in GRB 
and/or hypernova remnants, in their possible role as TeV unIDs.
There is evidence that typical GRBs predominantly occur in 
galaxies with less metals than our own \citep{fru06,sta06}, 
although the correlation between the host metallicity and the GRB 
energy is not confirmed \citep{sav08a,sav08b}. On the other hand broad-lined 
SNe Ic also inhabit more metal-rich galaxies \citep{mod08},
suggesting that at least hypernovae and possibly GRB/SNe with 
LL jets can occur in our Galaxy.
In addition, the evidence that GRBs occur in low metal regions could be biased
because the GRBs in the metal rich region tend to have no optical afterglows
due to the dust absorption (i.e., dark GRBs, which is about half of all GRBs).
Then it becomes difficult to identify the hosts for measuring the metal abundance.

To discriminate amongst these models, the imaging of the 
gamma-ray morphology would be useful.
The $\pi^0$ decay model predicts a shell structure,
since the density inside the remnants is low,
while the $\beta$ decay model predicts an elongated structure,
and the RI decay model predicts a center-filled structure.
For remnant ages $t_{\rm age}< 10^5$ yr, the SNR may be observed also
at other wavelengths.  HESS J1731-347 may be such an example,
which is likely to be an old SNR with a large gamma-ray to radio
flux ratio $F_{\rm TeV}/F_{\rm radio}\sim 33$ \citep{tian08}.
HESS J1834-087 may be also an old remnant 
%\citep{tian07}
with an elongated gamma-ray emission
outside the SNR \citep{aha06},
which could be interpreted in the conetxt of a
$\beta$-decay model (but see also \citet{muk08}).
%It will be important to observe
%possible GRB remnants such as SNR W49B.
%With either a detection or a non-detection, we can constrain the CR 
%acceleration mechanism in GRBs/hypernovae, the presence of a jet 
%component, and the GRB/hypernova rate in our Galaxy.

The $\pi^0$ decay model predicts detectable GeV emission
if the spectral index of $2.1$-$2.4$ continues down to the GeV region
(see Fig.~\ref{fig:flux}).
Other models predict relatively low levels of GeV emission
\citep[Fig.~\ref{fig:flux};][]{ioka04,yam06,atoyan06},
although rare young remnants may be bright if
$\varepsilon_p \sim 20 {\rm GeV} (t/10^4 {\rm yr})$
in the RI model. This may be tested by the Fermi satellite.
Note that the GRB prompt and afterglow emission will mask
the very early RI decay emission, which becomes prominent later.

The implied TeV neutrino flux ($\sim$ gamma-ray flux)
is not enough for detection by current facilities. However,
in the future one may in principle test models through the 
flavor ratio.  We expect $\pi^{+} \to \mu^{+} + \nu_{\mu}
\to e^+ + \nu_e + \bar \nu_{\mu} + \nu_{\mu}$ in the $\pi^0$ decay model,
no neutrinos in the $\beta$ decay model
(since neutrinos are beamed and usually off-axis), 
and ${}^{56}{\rm Co} \to {}^{56}{\rm Fe}^{*} + e^{+} + \nu_e$
in the RI decay model.

As discussed at the end of \S~\ref{sec:pi},
the radio observations limit the $(e/p)_{10{\rm GeV}}$ ratio to
$\lesssim 10^{-3}$ in the $\pi^0$ decay model.
For the RI decay model,
the CRs need to freely escape from the acceleration site
before suffering the adiabatic energy losses
(see the end of \S~\ref{sec:ri}).

The implied CR energy budget is less than $\sim 10\%
\sim (10\ {\rm times}\ {\rm energy}) \times (10^{-2}\ {\rm times}\ 
{\rm rate})$ of the standard SN CR energy budget in our models.
The CRs above the knee at $\sim 3 \times 10^{15}$ eV, however,
could be produced mainly by extragalactic and/or Galactic
GRBs/hypernovae \citep{wic04,minn06,wan07,bud08}.
The chemical composition is increasingly richer in heavy nuclei
above the knee $\sim 3 \times 10^{15}$ eV to the second knee
$\sim 6 \times 10^{17}$ eV \citep{ant05,abb05},
and possibly above $\sim 3 \times 10^{19}$ eV \citep{ung07}.
These may be accelerated by jets as in the RI decay model.

The GRB remnants may be also responsible for the excesses of cosmic-ray
positrons and electrons recently observed by the PAMELA 
and ATIC/PPB-BETS experiments \citep{ioka08,Fujita:2009wk}.
Since the electron and positron sources should be nearby 
(less than $\sim 1$kpc away),
it may be difficult to detect these sources by gamma-rays.

Recently, wide-field optical surveys have been discovering
exceptionally luminous SNe, such as SN 2005ap, SN 2008am, SN 2006gy, 
SN 2006tf and SN 2008es \citep{mil08}.
The SN 2008es yields a radiated energy of $\simg 10^{51}$ erg
and possibly a total energy of $\sim 10^{52}$ erg,
comparable to the hypernova energy.
The rate could be also comparable to the hypernovae rate
although only an upper limit to the rate is obtained \citep{mil08}.
Therefore such luminous SNe could also become TeV unIDs like the hypernova case,
although the luminous SNe are type II and may not be related with GRBs.

In conclusion, we propose that GRB/hypernova remnants are
promising candidates for the TeV unIDs.
If TeV unIDs are hypernova remnants,
we can predict more extended (and more numerous) TeV SNRs than observed,
which may be discovered by expanding the search region to larger
angular scales, even with the current instruments.
We also propose a new process of TeV gamma-ray emission
involving the decay of accelerated radioactive isotopes, such 
as ${}^{56}$Co entrained by relativistic
or semi-relativistic jets in GRBs/hypernovae.

\acknowledgments
We thank T.~Nakamura, T.~Kamae, S.~Razzaque, S.~Inoue, R.~Mukherjee, C.~Dermer,
R.~Yamazaki, S.~Park and T.~Mizuno for useful comments. 
This work is supported in part 
%by Grant-in-Aid for the 21st Century COE
%``Center for Diversity and Universality in Physics''
%from the Ministry of Education, Culture, Sports, Science and Technology
%(MEXT) of Japan and 
by the Grant-in-Aid from the 
Ministry of Education, Culture, Sports, Science and Technology
(MEXT) of Japan, No.18740147, 19047004, 21684014 (K.I.), 
and NASA NNX08AL40G (P.M.).

%\appendix

%\clearpage

%% the file, uncomment the \plotone, \plottwo, and
%% If you need a layout that cannot be achieved with \plotone or
%% \plottwo, you can invoke the graphicx package directly with the
%% \includegraphics command or use \plotfiddle. For more information,
%% please see the tutorial on "Using Electronic Art with AASTeX" in the
%% documentation section at the AASTeX Web site,
%% http://www.journals.uchicago.edu/AAS/AASTeX.
%%
%% The examples below also include sample markup for submission of
%% supplemental electronic materials. As always, be sure to check
%% the instructions to authors for the journal you are submitting to
%% for specific submissions guidelines as they vary from

%% This example uses \plotone to include an EPS file scaled to
%% 80% of its natural size with \epsscale. Its caption
%% has been written to indicate that additional figure parts will be
%% available in the electronic journal.

%\begin{figure}[t]
%\includegraphics[width=\linewidth]{Fig1}% Here is how to import EPS art
%\caption{}
%\end{figure}

%\clearpage

%\begin{figure}
%\plotone{f1.eps}
%\caption{flux
%}
%\label{fig:flux}
%\end{figure}

%\begin{figure}
%\plotone{f2.eps}
%\caption{
%}
%\label{fig:spec}
%\end{figure}

\end{document}